\documentclass[12pt,twoside]{article}   
\usepackage[super,sort,comma]{natbib}

\usepackage{fancyhdr}		

\newcommand{\captionv}[3]{\begin{center}\parbox{#1cm}{\caption[#2]{{\sf #3}}}
        \end{center}}

\usepackage[section]{placeins}   %

\usepackage{graphicx}

\makeatletter \renewcommand\@biblabel[1]{$^{#1}$} \makeatother
\newcommand{\cen}[1]{\begin{center} #1 \end{center}}

\usepackage[mathlines]{lineno}

\usepackage{hyperref}
\hypersetup{ colorlinks,
    citecolor=blue,
    filecolor=blue,
    linkcolor=blue,
    urlcolor=blue
}

\usepackage{xcolor}

\definecolor{gray}{rgb}{0.6,0.6,0.6}
\definecolor{red}{rgb}{0.85,0,0}
\definecolor{green}{rgb}{0,0.85,0}
\definecolor{blue}{rgb}{0,0,0.85}
\definecolor{beige}{rgb}{0.92,0.87,0.78}
\usepackage[all]{hypcap}    

\begin{document}

\cen{\sf {\Large {\bfseries Prediction of post-radiotherapy recurrence volumes in head and neck squamous cell carcinoma using 3D U-Net segmentation} \\  
\vspace*{10mm}
Denis Kutnár$ ^{1,2}$, Ivan R Vogelius$ ^{2,3}$, Katrin Elisabet Håkansson$ ^{2}$, Jens Petersen$ ^{1,2}$, Jeppe Friborg$ ^{2}$, Lena Specht$ ^{2,3}$, Mogens Bernsdorf$ ^{2}$, Anita Gothelf$ ^{2}$, Claus Kristensen$ ^{2}$, Abraham George Smith$ ^{1,2}$} \\
1 - Dept. of Computer Science, University of Copenhagen, Copenhagen, Denmark; \\
2 - Dept. of Oncology, Copenhagen, Copenhagen University Hospital – Rigshospitalet, Copenhagen, Denmark
 \\
3 - Dept. of Health and Medical Sciences, University of Copenhagen, Copenhagen, Denmark 
\vspace{5mm}\\
Version typeset \today\\
}

\pagenumbering{roman}
\setcounter{page}{1}
\pagestyle{plain}
denis.kutnar@regionh.dk\\

\begin{abstract}
\noindent {\bf Background:} Locoregional recurrences (LRR) are still a frequent site of treatment failure for head and neck squamous cell carcinoma (HNSCC) patients.
{\bf Aim:} Identification of high risk subvolumes based on pretreatment imaging is key to biologically targeted radiation therapy. We investigated the extent to which a Convolutional neural network (CNN) is able to predict LRR volumes based on pre-treatment $ ^{18}$F-fluorodeoxyglucose positron emission tomography (FDG-PET)/computed tomography (CT) scans in HNSCC patients and thus the potential to identify biological high risk volumes using CNNs. \\ 
{\bf Methods:} For 37 patients who had undergone primary radiotherapy for oropharyngeal squamous cell carcinoma, five oncologists contoured the relapse volumes on recurrence CT scans. Datasets of pre-treatment FDG-PET/CT, gross tumour volume (GTV) and contoured relapse for each of the patients were randomly divided into training (n=23), validation (n=7) and test (n=7) datasets. We compared a CNN trained from scratch, a pre-trained CNN, a SUVmax threshold approach, and using the GTV directly.
{\bf Results:} The SUVmax threshold method included 5 out of the 7 relapse origin points within a volume of median 4.6 cubic centimetres (cc). Both the GTV contour and best CNN segmentations included the relapse origin 6 out of 7 times with median volumes of 28 and 18 cc respectively. \\
{\bf Conclusion:} The CNN included the same or greater number of relapse volume POs, with  significantly smaller relapse volumes. Our novel findings indicate that CNNs may predict LRR, yet further work on dataset development is required to attain clinically useful prediction accuracy. \\

\end{abstract}




\setlength{\baselineskip}{0.7cm}      

\pagenumbering{arabic}
\setcounter{page}{1}
\pagestyle{fancy}

\section{Introduction}
Radiotherapy for head and neck squamous cell carcinoma (HNSCC) is associated with a substantial burden of toxicity, despite substantial improvements in treatment conformity and delivery. Patients still suffer a relatively high rate of locoregional recurrence (LRR), despite being treated to tolerance, where further uniform dose-escalation seems infeasible. Already in year 2000 Ling et al. \cite{ling2000towards} suggested using biological imaging to define a \textit{biological target volume} that may be at a particular high risk of relapse and in 2005 the concept of \textit{dose painting by numbers} \cite{bentzen2005theragnostic} was suggested to use theragnostic imaging to tailor a nonuniform radiation dose to the risk profile of intratumour heterogeneity. These concepts are ideally suited for HNSCC to improve treatment response without the excess toxicity associated with uniform dose escalation.

A small number of early phase clinical trials have tested these concepts in practice \cite{rasmussen2016phase,  welz2022dose, duprez2011adaptive}, but the toxicity burden is non-trivial \cite{rasmussen2016phase, olteanu2018late} and no large randomized trials have proven the value of the approach in clinical practice. A substantial challenge is to identify the right volume to boost - and a sufficiently small volume to limit the burden of treatment. Prior work has focused primarily on hypoxia imaging or $ ^{18}$Fluorodeoxyglucose (FDG).

One key bottleneck is identifying subregions  with demonstrable higher risk disease. Associations between multimodal imaging and pathological features of the tumour appear less straightforward in clinical samples than textbook explanations \cite{rasmussen2020does, rasmussen2021intratumor}. Thus a purely data driven approach to identify subregions of tumour harboring most recurrences based on pretreatement or longitudinal FDG positron emission tomography (PET)/computed tomography (CT) have been attempted \cite{berwouts2013three, rasmussen2016phase, due2014recurrences}. Ideally, the method to predict regions of high risk of recurrence should be flexible enough to accommodate multiple imaging modalities and potentially also supporting clinical data.

Convolutional neural networks (CNNs), a subset of deep learning, have demonstrated increasingly significant promise over the past five years across a wide range of medical domains\cite{isensee2018nnu, heller2019kits19, wang2017searching, icsin2016review}, outperforming more conventional techniques. CNNs play a significant part in a number of medical processes, including tumour and organs at risk segmentation, diagnosis, and treatment planning.

To the best of our knowledge, no prior work has investigated the potential of CNNs to predict LRR at the voxel level. In light of this, the aim of this study was to investigate the extent to which a CNN is able to segment LRR from pre-treatment FDG-PET/CT images.


\section{Materials and Methods}

\begin{figure}[ht]
   \begin{center}
   \includegraphics[width=16cm]{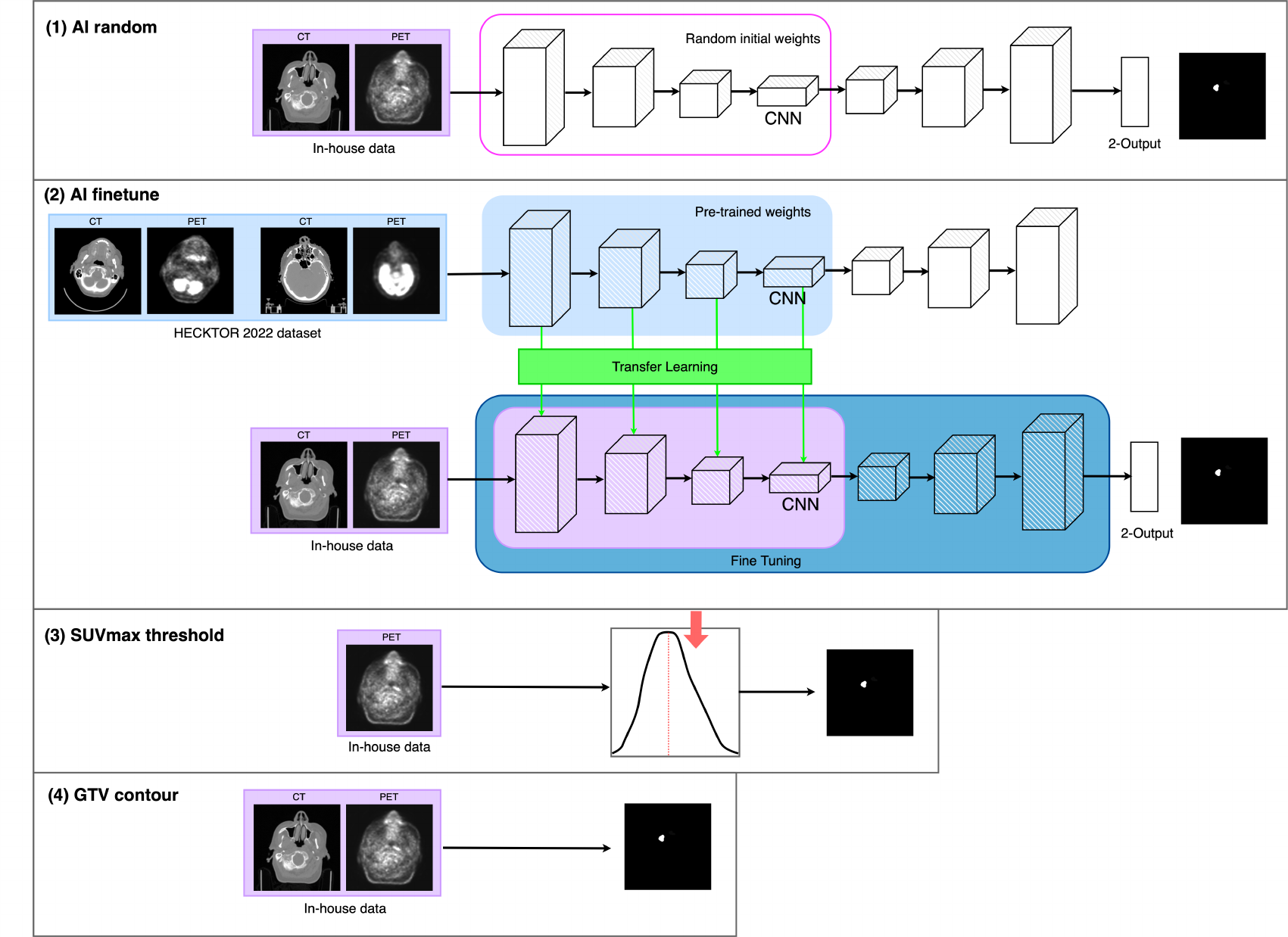}
   \captionv{14}{Short title - can be blank}
   {Schematic illustration of four distinct model-building approaches tested to identify the voxel-level locations from pre-treatment FDG-PET/CT scans where locoregional recurrences are most likely to develop. The top row (1) ``AI random" represents a CNN trained from random initial weights. The second row (2) ``AI finetune", is a CNN firstly pre-trained on a larger dataset. The third row (3) ``SUVmax threshold” illustrates the approach using thresholding based on SUVmax and the last row (4) ``GTV contour”, is a method using the GTV contour directly as the recurrence prediction.
   \label{figure1} 
    }  
    \end{center}
\end{figure}

We compared four distinct methods to predict sub-regions from FDG-PET/CT images where LRR is most likely to appear, as shown in Figure \ref{figure1}. (1) ``AI random", which represents a CNN trained from random initial weights, (2) ``AI finetune", which defines the same model but is first pre-trained on a larger dataset, (3) ``SUVmax threshold", a method where thresholding was performed based on the maximum standardised uptake value (SUVmax), and (4) ``GTV contour", a method that used the gross tumour volume (GTV) directly as a recurrence prediction.

\vspace{0.5cm}

\subsection{Data and pre-processing}
This study employed two separate datasets: data from the publicly available third edition of the HEad and neCK TumOR (HECKTOR2022) challenge dataset and in-house data from patients with squamous cell carcinoma of the oropharynx (OPSCC) treated with definitive radiation therapy at Rigshospitalet, Copenhagen, Denmark. 

The HECKTOR2022 challenge dataset consists of patients with histologically confirmed oropharyngeal head and neck cancer who received radiation with or without chemotherapy \cite{andrearczyk2022overview}. A total of 524 PET/CT images, including low-dose, non-contrast-enhanced CT and registered FDG-PET images, were gathered from 9 different institutions. In addition to the image data, each patient's clinical information is made available, including the gender, age, weight, alcohol and tobacco abuse, human papilloma virus status, and type of therapy (radiotherapy only or chemoradiotherapy). This cohort (21\% of recurrence occurrences and a median recurrence-free survival of 14 months) was used to represent a reasonable approximation of the distribution of the real-world patient population referred for radiotherapy.

The HECKTOR dataset is available through the challenge website\footnote{ \href{https://hecktor.grand-challenge.org/Data/}{https://hecktor.grand-challenge.org/Data/}}.

\begin{figure}[ht]
   \begin{center}
   \includegraphics[width=8cm]{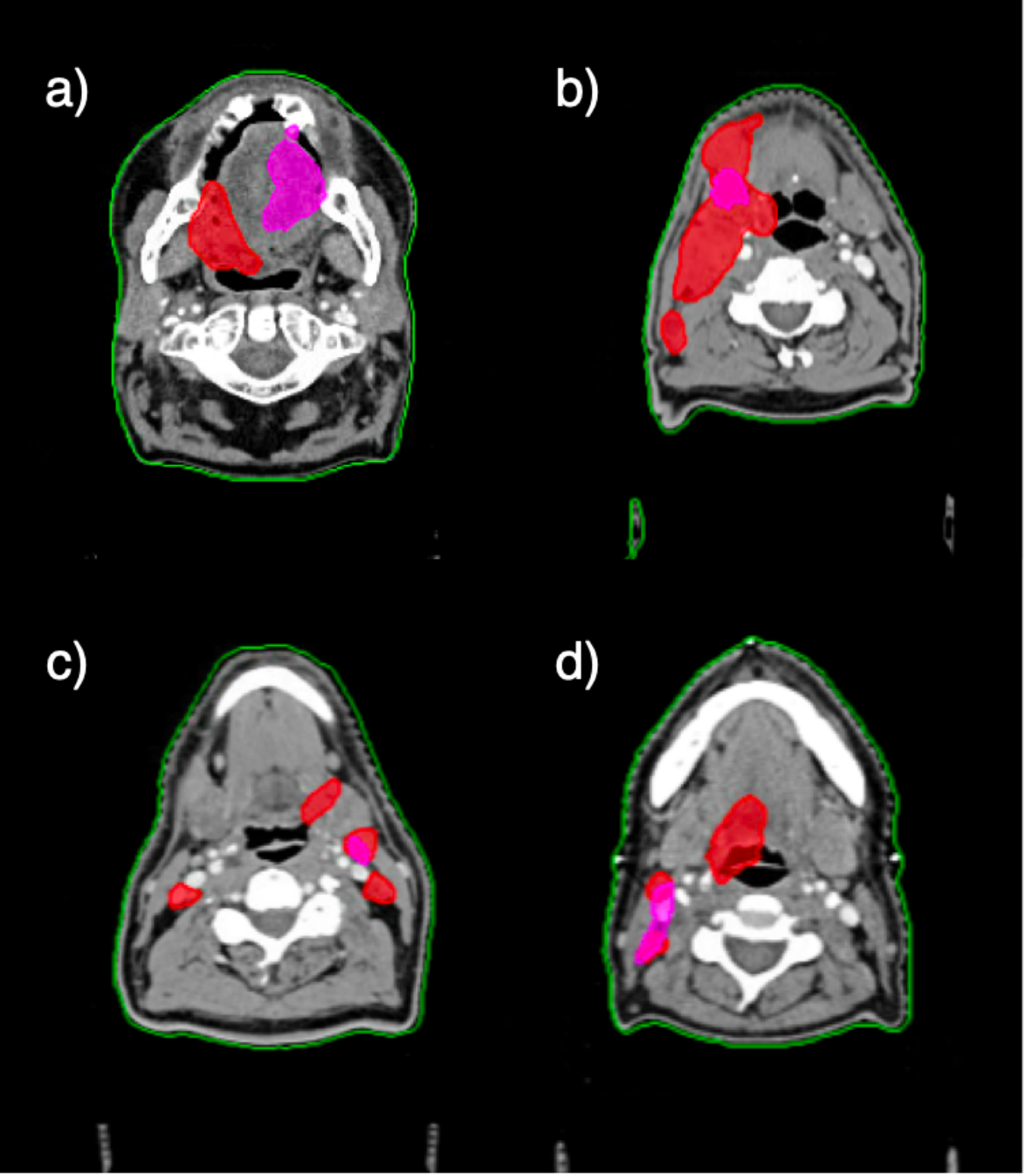}
   \captionv{14}{Short title - can be blank}
   {Example of four randomly picked patients with head and neck squamous cell carcinomas included in the Rigshospitalet study. The figure shows one slice for each patient (a-d) in the transverse plane of the planning CT. Red contour: pre-treatment GTV. Pink contour: Relapse.
   \label{figure2} 
    }  
    \end{center}
\end{figure}

For the recurrence prediction, we retrospectively collected 46 patients who had undergone primary radiotherapy for OPSCC at Rigshospitalet, Denmark, between 2009 and 2017.  All patients included in the study developed isolated loco-regional relapse between three months and five years after the completion of the treatment, verified by biopsy\cite{katrin}, as shown in Figure \ref{figure2}. A total dose of 66 or 68 Gy (2 Gy/fraction) using volumetric modulated arc therapy or intensity-modulated radiation therapy was administered to the patients in accordance with the Danish Head and Neck Cancer Group (DAHANCA) guidelines\cite{overgaard2016danish}. The collection consisted of rigidly registered pre-treatment FDG-PET/CT and relapse CT, PET/CT, or magnetic resonance scans with a median follow-up of 1.3 years. 

The pre-processing involved the following steps. First, we anonymised data using the Eclipse Treatment Planning System (TPS) (Varian Medical Systems UK). Then the pre-treatment FDG-PET/CT and relapse images were exported to Digital Imaging and Communications in Medicine (DICOM) file format and checked for consistency across the imaging modalities. In cases where one of the modalities was available exclusively in the head and neck (H\&N) region rather than a full-body scan, the pre-treatment PET or CT with a corresponding relapse image were cropped in the transverse plane, according to the physical slice location determined in DICOM metadata. The voxel spacing was uniform for both modalities, with a spacing of 0.9765 x  0.9765 x 2.0  for CT and 2.036 x 2.036 x 2.0 for PET images. We resampled PET  to CT using the SimpleITK \cite{yaniv2018simpleitk} linear interpolation resample function to match the voxel size, voxel spacing, image origin, and image orientation. The image normalization strategy was inspired by the winning solution of the MICCAI 2020 HECKTOR segmentation challenge\cite{iantsen2020squeeze}. Accordingly, CT intensities were clipped in the range of [-1024, 1024] Hounsfield Units and then mapped to [-1, 1]. PET images were normalised using the Z-score normalisation technique performed on a patient level. 

The majority of the images used in our study were full-body scans. However, we were solely interested in the H\&N region where the GTV and relapses are present. To crop the images, we determined the maximum distance of the GTV’s centre and its furthest point of the fused GTV and relapse volume across the entire training and validation set. Additionally, we added 15\% of padding to improve generalizability since this estimate was in-house training data-specific. Starting from the centre of the GTV, all images were cropped using a 3D bounding box with dimensions of 160 x 224 x 128 in the X-Y-Z coordinate system.

\subsection{Ground truth}
The dataset from Rigshospitalet incorporated multiple radiotherapy target volumes defined based on the pre-treatment FDG-PET/CT images. The annotation of the GTV was performed by a radiologist and an oncologist, in collaboration. Five oncologists participated in the annotation process of relapse volumes using the Eclipse TPS. Patients' information, including medical records and images, was available prior to annotation. A total of sixty relapse volumes were identified in the 46 patients.

\begin{figure}[ht]
   \begin{center}
   \includegraphics[width=10cm]{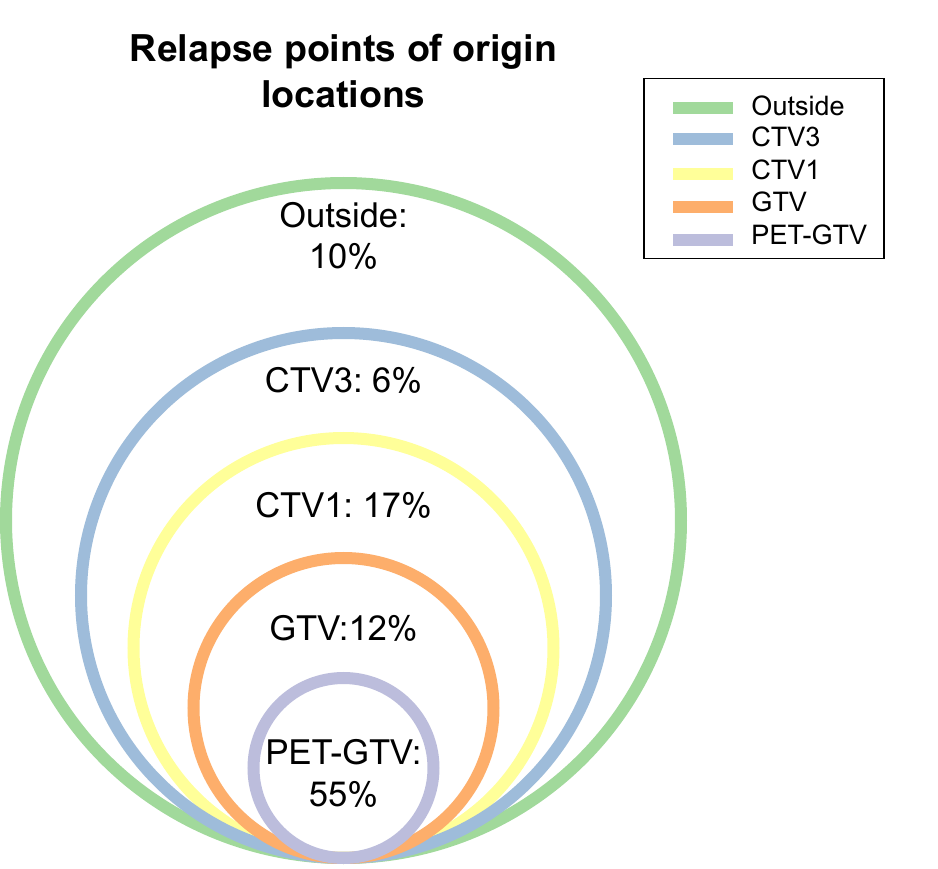}
   \captionv{14}{Short title - can be blank}
   {The distribution of the sites where the relapse volumes originated for 46 patients included in the Rigshospitalet study. In purple: PET-GTV with 55\%. In the orange: GTV (excluding the PET-GTV) with 12\%. In yellow, CTV1 (excluding the GTV) with 17\%. In blue: CTV3 (excluding the CTV2) with 6\%. None of the relapses originated specifically in CTV2 and 10\% developed outside the target volumes, marked in green 
\cite{katrin}.
   \label{figure3} 
    }  
    \end{center}
\end{figure}

As described by Håkansson et al.\cite{katrin}, 55\% of points of origin (PO) were encompassed by the PET-GTV (FDG-PET avid gross tumour volume), followed by 12\% in the GTV excluding the PET-GTV, 17\% in the high-risk clinical target volume (CTV1) excluding the GTV, and 6\% in the low-risk clinical target volume (CTV3) excluding the intermediate risk clinical target volume (CTV2). Ten percent of relapses originated outside the target volumes as illustrated in Figure \ref{figure3}.

The relapse volume PO was estimated as the centre of volume\cite{katrin, due2010methodologies, zukauskaite2017analysis}. A patient may develop one or multiple relapse volumes. Thus, we performed a connected component analysis to identify the individual volumes per patient. Consequently, to define the POs, we applied the morphological binary erosion operator with a structuring element of 3x3x3 for each component found.

\subsection{Data partition}
The subjects were randomly divided into training, validation and testing groups at a ratio of 6:2:2. The testing group remained unseen throughout all development and experimentation phases and was only utilised at the end to compare and evaluate the methods.

\subsection{Model architecture}
We adopted a 3D variant \cite{cciccek20163d} of the U-Net\cite{ronneberger2015u} architecture designed for biomedical image segmentation, freely available on GitHub \cite{cciccek20163d}. We adjusted the 3D U-Net from the original implementation to take two-channel input, representing two imaging modalities, CT and PET and replaced batch normalisation with instance normalisation \cite{ulyanov2016instance},  which is more appropriate for small batch sizes \cite{wu2018group}. In summary, a total of 65 layers and more than 5.6 million trainable parameters formed the network, implemented using PyTorch\cite{paszke2019pytorch} version 1.12.

\subsection{Experiments settings}
Initial experiments were done exclusively on the training and validation subsets and were conducted in the same environment and device configuration, using two NVIDIA GeForce RTX 3090 GPUs and an AMD Ryzen Threadripper 3960X CPU. The network depth, optimizer, and learning rate, as well as the best performing CNN for the (1) AI random and (2) AI finetune methods, were determined using the training and validation datasets.

\subsection{Training strategies}

\textbf{AI random:}\\

We trained five CNN models from random initial weights using the He initialization method\cite{he2015delving}. The He initialization initializes the weights using a Gaussian distribution with zero mean and a variance of 2/n, where n is the number of input neurons in the layer. This initialization method has been found to work well for deep neural networks and has been widely used in various computer vision tasks\cite{he2019rethinking}.

For optimization, we employed the Adam optimizer \cite{zhang2018improved} with the following additional parameters: beta1 = 0.9, beta2 = 0.999, and epsilon = 1e-8. The Adam optimizer has proven to be successful and popular in training deep learning models. We used a learning rate of 0.1 for the initial training. The learning rate was then reduced by a factor of 0.05 following ten iterations in which the metrics remained unchanged.

Furthermore, to address the class imbalance between the background and foreground (relapse volume), we utilized the Dice loss function, adopted from the MONAI open-source framework \cite{cardoso2022monai}. Dice loss is commonly used for segmentation tasks and has shown effectiveness in handling class imbalances\cite{sudre2017generalised}. We set the smoothing parameter of the Dice loss to 1e-05.

In summary, the model required a five-dimensional input that can be represented as a tensor of dimension [2, 2, 128, 224, 160], where the first dimension specifies the batch size, the second is the number of input channels (PET/CT), the third is the patch size along the z-axes, and the final two dimensions are height and width. The model outputs a 3D probability map of the relapse volume. 

We used early-stopping, with training ceasing after 60 epochs without improvement in the evaluation metric measured on the validation set. To reduce overfitting, we augmented the training set by performing a random horizontal flip with a probability of 10\%. After all five models had been trained, the model with the highest value in the evaluation metric calculated on the validation set was chosen for inference on the test set.\par
\noindent
\textbf{AI finetune:}\\
The Rigshospitalet dataset is a relatively small cohort in terms of CNN model training. We hypothesized that initial pre-training on a similar domain might help to achieve a better result. As a first step, we pre-trained the model on the HECKTOR2022 dataset and stored the model weights. We then trained 5 relapse segmentation CNN models initialised from the pre-trained weights by adhering to the same hyperparameters and training strategies as those described in the AI Random model section.\par

\noindent
\textbf{SUVmax threshold:}\\
SUVmax of FDG-PET may play an important role in the prediction of local recurrence\cite{takeda2011maximum, vogelius2013failure}. We performed SUVmax thresholding on pre-treatment PET images to  estimate the area where the relapse volume originated. The highest voxel value found within the GTV formed the basis for the SUVmax\cite{paidpally2012fdg}. The PET images were then thresholded using a range of percentiles calculated from the SUVmax, starting from 1\% - 100\%, with a step size of 1, as shown in Figure \ref{figure4}. Then, we excluded every voxel contained in a brain mask from the thresholded estimate. \par

\noindent
\textbf{GTV contour:}\\
As a simple baseline method we also directly used the GTV contour as the relapse volume prediction, since more than half of the relapses originated in the GTV. 

\subsection{Evaluation}
The metrics were calculated between the contoured relapse volumes (ground truth) and the predicted volumes for each method. We reported recall, precision, and Sørensen-Dice similarity coefficient (Dice) for every approach (computed with MedPy library \cite{Maier}). Dice coefficient is a frequently employed evaluation metric to quantify the performance of medical image segmentation methods\cite{bertels2019optimizing}. For each method, we also determined the size of the predicted volumes in cubic centimetres (cc) and how many of the relapse POs were captured.

\subsection{Methods comparison}

We performed a paired t-test to compare the AI random  and AI finetune models in terms of Dice coefficient on the test set. Similarly, we compared the best AI approach to the SUVmax and GTV contour baseline methods in terms of Dice coefficient and predicted relapse volumes. 
We computed a Fisher exact test to compare the best CNN to the SUVmax and GTV contour baseline methods in terms of the number of ground truth POs included in the predicted relapse regions.

\section{Results}

\subsection{Data}
For the Rigshospitalet data the PET scanning could not be completed for three patients due to interfering medical conditions. A relapse contour could not be exported for two patients, or was not present. A PET scan for one patient was corrupted. Three patients either exhibited uneven voxel spacing in the transverse plane or missed DICOM slices. A total of 37 individuals were enrolled in the research after these nine patients were excluded. The training, validation and testing groups included 23, 7, and 7 patients, respectively.

\subsection{Training}

\begin{figure}[ht]
   \begin{center}
   \includegraphics[width=10cm]{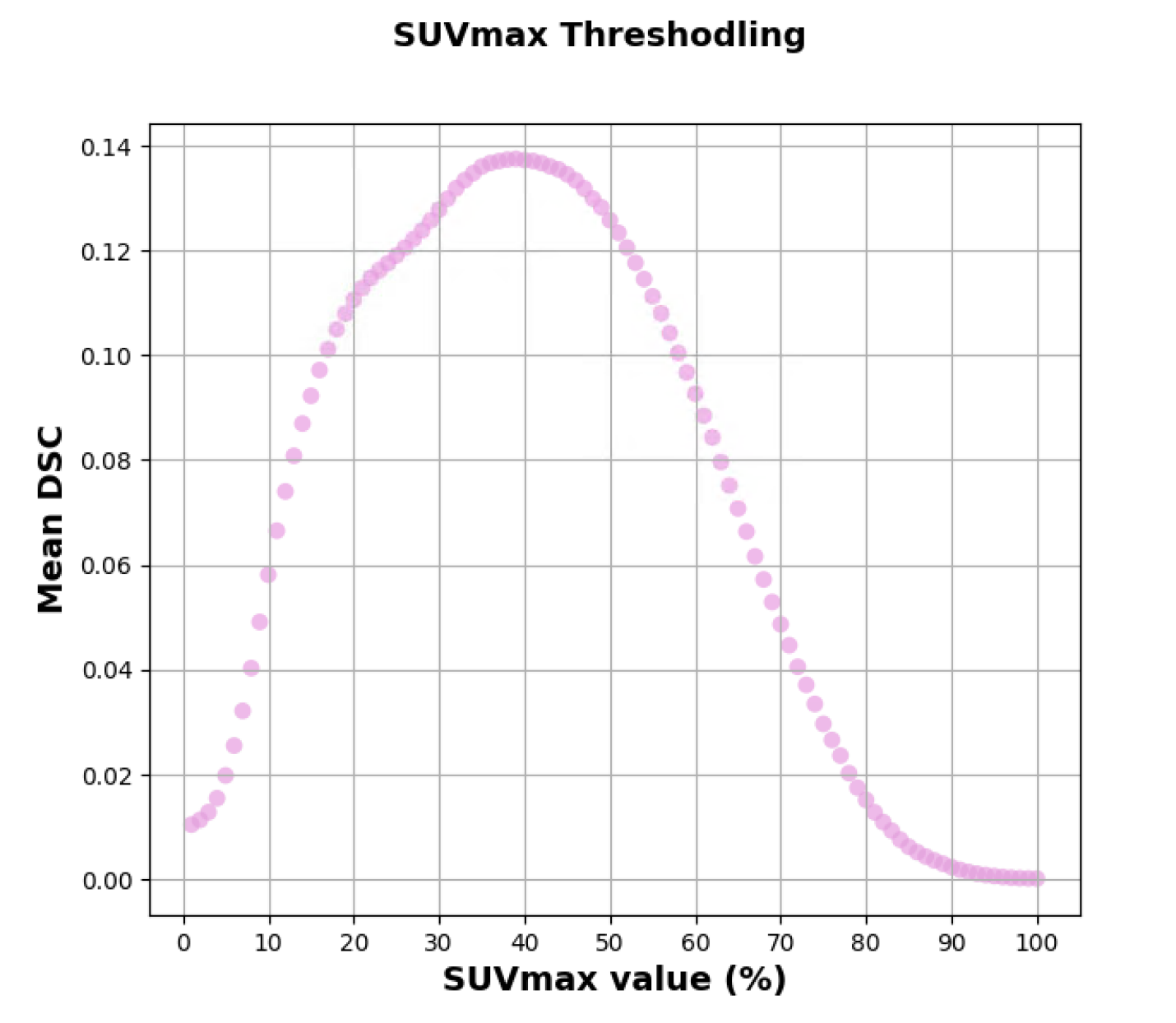}
   \captionv{14}{Short title - can be blank}
   {Scatter plot showing different Dice coefficients obtained for a range (1\% - 100\%) percentiles using SUVmax threshold method. The best Dice coefficient was obtained when taking 38\% of the maximum standardised uptake value from the selected region of interest. 
   \label{figure4} 
    }  
    \end{center}
\end{figure}

The best AI random model had a mean validation set Dice coefficient of 0.54, which was found on epoch 66. Each epoch, including training and validation, took an average of 4.26 minutes. 
The tumour segmentation model, which was trained using the HECKTOR2022 dataset, had a mean Dice coefficient of 0.75 on the HECKTOR2022 validation dataset.
The relapse segmentation training process of the fine-tuned model resulted in a mean Dice coefficient of 0.16 on the validation set, found on epoch 19.
We found that 38 was the ideal percentage for the SUVmax after doing 100 trials (Figure \ref{figure4}), one for each percentage, producing a mean Dice coefficient of 0.14 on the validation set. Consequently, we thresholded the test images with 38\% of the SUVmax.

\subsection{Test set evaluation}

For each method we report the median Dice coefficient with Interquartile Range (IQR), predicted volume size with IQR and the number of predicted relapse volumes containing the PO (Table \ref{table1}).  A short comparison of the best CNN approach (selected based on validation set Dice coefficient) compared to the SUVmax and GTV contour baseline methods can be found in Table \ref{table1}. The highest median Dice coefficient of 0.29 with (0.24, 0.64) IQR was given by the GTV contour. 

\begin{figure}[ht]
   \begin{center}
   \includegraphics[width=10cm]{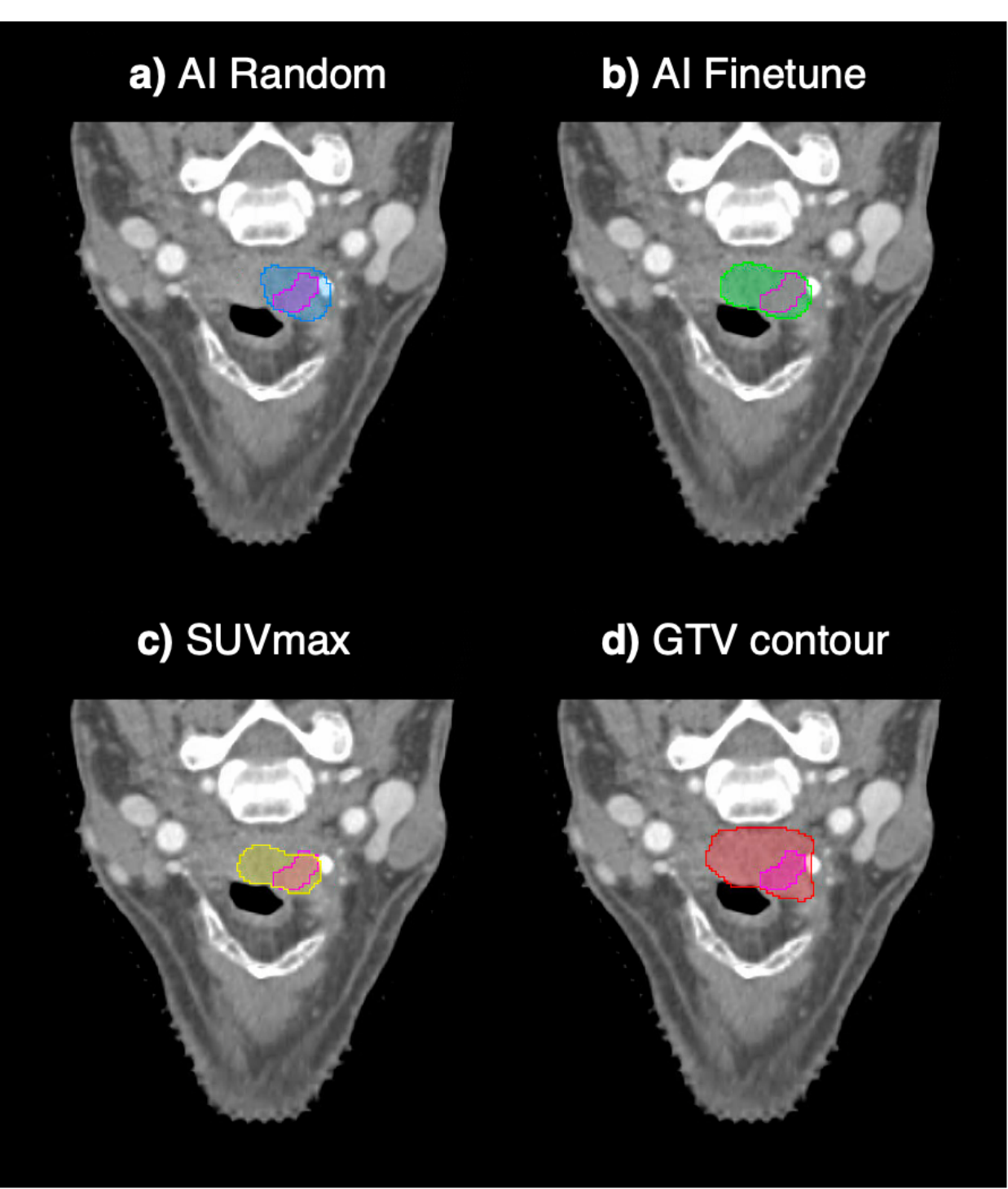}
   \captionv{14}{Short title - can be blank}
   {Visual assessment and comparison of the performance of four distinct methods computed on the test set for post-radiotherapy recurrence prediction. The figure shows a craniocaudal CT slice of patient five from the test group, suffering from head and neck squamous cell carcinomas, who developed one relapse volume 1.1 years after completing treatment. Top-left image (a) shows the relapse volume (ground truth) in pink and the AI Random method’s prediction in blue. Top-right image (b) pink contour is the relapse volume, green contour is the prediction from the AI Finetune model. Bottom-left (c) shows the ground truth in pink and the estimated relapse volume obtained using the SUVmax threshold method (38\% of maximum standardised uptake value) in yellow. Bottom-right (d) GTV contour used directly as the recurrence prediction indicated in red and the ground truth in pink.
   \label{figure5} 
    }  
    \end{center}
\end{figure}

We did not find a significant difference between the AI random and AI finetune models when comparing Dice coefficient (p $>$ 0.29), even though the AI random model tended to produce more accurate results. In Figure \ref{figure5}a, we show an example of recurrence segmentation using the AI random model.

SUVmax thresholding included five out of the seven relapse origin points within a median volume of 4.6 cc  (Table \ref{table1}). Both the GTV Contour and CNN (AI random) segmentations included the relapse origin six out of seven times with median volumes of 28 and 18 cc respectively (Table \ref{table1}). The CNN (AI random) relapse segmentation was significantly smaller in volume than the GTV contour baseline (p $<$ 0.018), yet still contained the PO the same number of times.

We found no statistically significant difference between the methods in terms of the number of ground truth POs included in the predicted relapse regions.

\vspace{0.5cm}

\begin{table}[ht]
   \begin{center}
   \includegraphics[width=12cm]{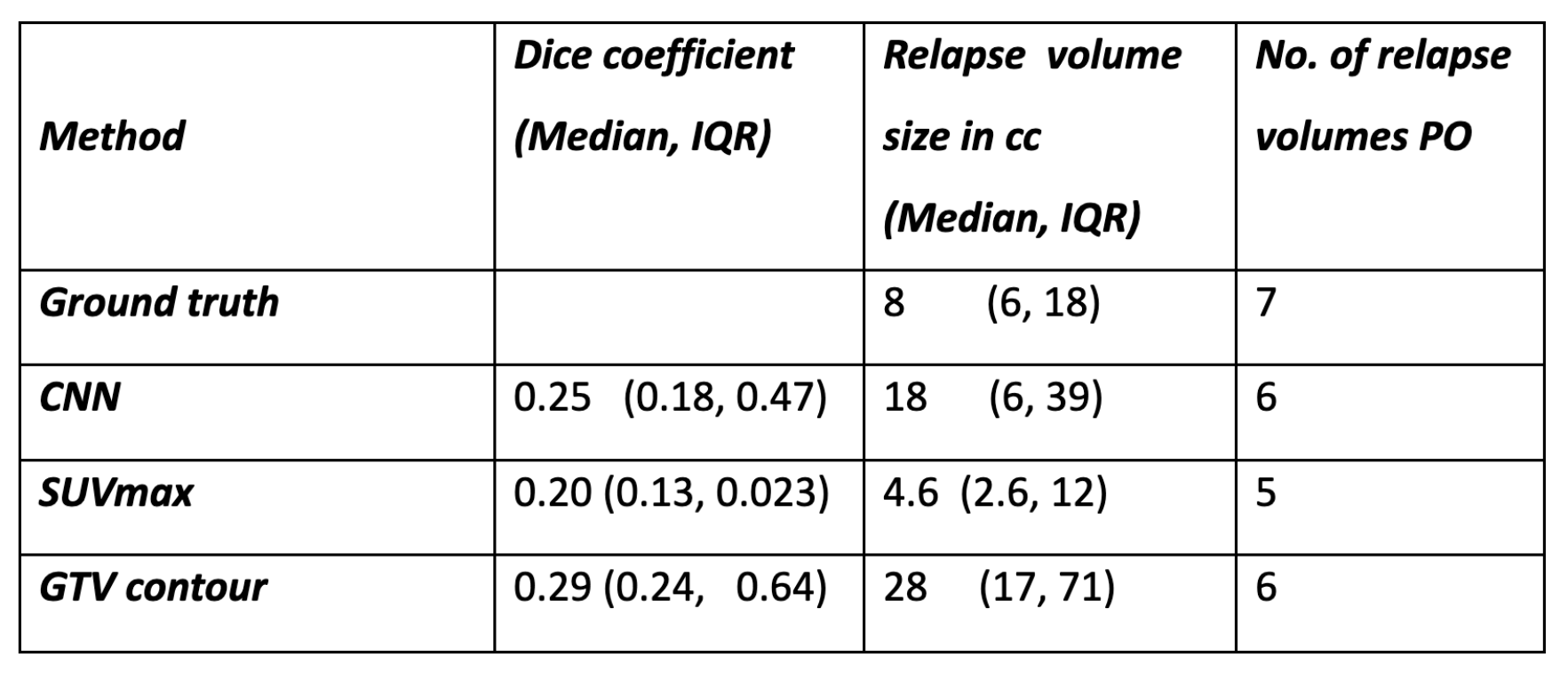}
   \captionv{14}{Short title - can be blank}
   {Comparison of the best AI approach to the SUVmax threshold and the GTV contour methods for their attained Dice coefficients, the size of the predicted relapse volumes (measured in cubic centimeters - cc), and the number of detected relapse volumes point of origins computed on the test dataset containing seven patients.  Abbreviation 'IQR' stands for the interquartile range.
   \label{table1} 
    }  
    \end{center}
\end{table}

\section{Discussion}
It should be acknowledged that the task of segmenting a later LRR based on only pre-treatment data is extremely challenging. The resulting quantitative metrics for the segmentation accuracy are therefore much lower than, for example, tumour delineation.  For comparison, the 1st place solution of the Multimodal Brain Tumor Segmentation Challenge (BraTS) 2019 achieved a Dice coefficient of 0.89 \cite{jiang2019two}. Nevertheless, the task is of interest due to the potential of assigning voxel-based risks and modifying the therapeutic dose accordingly \cite{ling2000towards,bentzen2005theragnostic,vogelius2013failure}

To the best of our knowledge, our study is the first to investigate the extent to which deep learning is able to segment LRR at the voxel level solely from pre-treatment FDG-PET/CT images in patients suffering from HNSCC. The majority of research studies on recurrence prediction mainly focus on classification (predicting whether the recurrence will appear in the future) rather than segmentation. A study by Beaumont et al.\cite{beaumont2019voxel} was the first analysis to compute the probability of a voxel belonging to a recurrence region within the GTV using Radiomics and the Random Forest machine learning algorithm. In our study, we investigated the potential of CNNs to segment the recurrence at the voxel level including regional recurrences. 

We trained two CNN models to segment LRR. The first model (AI Random) was trained from random initial weights and achieved a mean Dice coefficient of 0.26 on the hidden test group. 
A mean Dice coefficient of 0.16 was determined for the second model (AI Finetune), which was first pre-trained on the larger HECKTOR2022 dataset encompassing primary tumours as we hypothesized that such pre-training would be beneficial in light of the limited number of recurrences available for a deep learning model to be trained from scratch. We speculate that model pre-training was not beneficial for recurrence segmentation because the task of segmenting recurrence may not have high affinity \cite{fifty2021efficiently} with the task of segmenting tumours.

The CNN segmentation displayed promising prediction capabilities, including the same or a greater number of relapse volume POs, whilst providing a significantly smaller relapse volume than other methods. Despite the potential for defining subvolumes of GTV to boost in high-risk patients, due to the limited accuracy, we cannot yet conclude on the suitability for clinical use.

Our model accuracy  is limited by the small size of the clinical dataset, including only 37 PET/CT images in total, of which we used 21 for training.  Major successes of AI reported in literature often rely on massive datasets. For example, the ``RSNA 2019 Brain CT Haemorrhage Challenge" \cite{flanders2020construction} included 874,035 CT images available for CNN models to detect brain haemorrhages. Larger datasets with paired baseline and recurrence data would likely enable us to obtain improved results. However, a limitation to such dataset building is data sharing and contouring the relapse, which is not done as part of routine clinical practice and requires medical expertise. Nevertheless, it seems prudent, in the light of massive investments in high precision radiotherapy, to document the failures more accurately and systematically than in current practice, for example using collaborative DICOM databases \cite{westberg2014dicom}. If this situation is improved and data is collected, deep learning appears to be a candidate method to analyze the resulting image data.

It is unrealistic that any model of later recurrence prediction could avoid predicting false positives regions (i.e. regions outside the later recurrence region, cf. Figure 5). Therefore the Dice coefficient and the recall (sensitivity) metrics should be expected to be low compared to other segmentation tasks. The question is if the false positive regions are too large for a realistic boost dose to be applied and, of course, if a simpler and more transparent method exists. Here, the FDG positive volume has previously been tested clinically in early phase dose escalation trials \cite{rasmussen2016phase} and should be considered the reference method against which to compare the CNN approach. The results of Table 1 combined with the well-established understanding that CNNs deliver improved outcomes with increased training data suggest the plausibility of deep learning methods enhancing prediction capabilities. Furthermore, the CNN output offers another advantage in the form of an adjustable threshold, enabling the method's customization according to the desired balance between sensitivity and specificity, or even facilitating a framework for dose painting by numbers. Moreover, a notable strength of deep learning models lies in their potential to seamlessly integrate image data with clinical information in predictions. For instance, variables such as HPV status, dose distribution, and stage can be effortlessly incorporated into deep learning models, eliminating the need to explicitly program their hypothesized impact on recurrence volume prediction.

\section{Conclusion}
The aim of this work was to investigate the extent to which CNNs are able to predict LRR at the voxel-level solely from pre-treatment FDG-PET/CT scans in HNSCC patients. The qualitative and quantitative evaluation performed on the test dataset highlight the difficult nature of LRR segmentation using the employed methods. However, the CNN method indicated the potential to detect relapse volume PO whilst providing a significantly smaller relapse volume size than the GTV baseline method.
Nevertheless, due to the limited accuracy, we cannot yet conclude on the suitability for clinical use. Our novel results show that CNNs could possibly serve as a method for predicting LRR in the future, benefiting from larger datasets. Expanding data collection of recurrences  holds promise for further investigating and harnessing CNNs' predictive capabilities in identifying regions at high risk of harbouring LRR.

\section{Acknowledgments}
This work was supported by the Danish Cancer Society (grant no R269-A15840 and R231-A13976).

\section{Conflict of Interest Statement}
The authors declare no conflicts of interest.

\newpage     

\clearpage


\section*{References}
\addcontentsline{toc}{section}{\numberline{}References}
\vspace*{-20mm}





\bibliography{references}      



\bibliographystyle{medphy.bst}    


\end{document}